\documentclass[prb,preprint,showpacs,preprintnumbers,amsmath,amssymb]{revtex4}

\usepackage{graphicx}
\usepackage{bm}

\newcommand{\be}{\begin{equation}}
\newcommand{\ee}{\end{equation}}
\newcommand{\eq}[1]{Eq.~(\ref{#1})}
\newcommand{\fig}[1]{Fig.~\ref{#1}}
\def\bea{\begin{eqnarray}}
\def\eea{\end{eqnarray}}

\def\vq{{\bf q}}

\def\vk{{\bf k}}

\def\qp{{\bf q}_{\parallel}}

\begin{document}

\title{Plasmon excitations in layered high-$\boldsymbol{T_c}$ cuprates} 

\author{Andr\'es Greco$^\dag$, Hiroyuki Yamase$^\ddag$, and Mat\'{\i}as Bejas$^\dag$}
\affiliation{
{$^\dag$}Facultad de Ciencias Exactas, Ingenier\'{\i}a y Agrimensura and
Instituto de F\'{\i}sica Rosario (UNR-CONICET),
Av. Pellegrini 250, 2000 Rosario, Argentina\\		
{$^\ddag$}National Institute for Materials Science, Tsukuba 305-0047, Japan
}

\date{\today}

\begin{abstract}
Motivated by the recent resonant inelastic x-ray scattering (RIXS) experiment for 
the electron-doped cuprates Nd$_{2-x}$Ce$_{x}$CuO$_{4}$ with $x \approx 0.15$,  
we compute the density-density correlation function 
in the $t$-$J$ model on a square lattice by including interlayer hopping and 
the long-range Coulomb interaction. We find that collective charge excitations 
are realized not inside the particle-hole continuum, 
but above the continuum as plasmons. 
The plasmon mode has a rather flat dispersion near the in-plane momentum $\qp=(0,0)$ 
with a typical excitation energy
of the order of the intralayer hopping $t$
when the out-of-plane momentum $q_{z}$ is zero. 
However, when $q_{z}$ becomes finite, the plasmon dispersion changes drastically 
near $\qp=(0,0)$, leading to a strong dispersive feature with an excitation gap scaled by 
the interlayer hopping $t_z$. We discuss the mode recently observed by RIXS near $\qp=(0,0)$ 
in terms of the plasmon mode with a finite $q_{z}$. 
\end{abstract}

\maketitle
\section{introduction}
Recently x-ray scattering revealed charge excitations for various hole-doped cuprates such as 
Y- \cite{ghiringhelli12, chang12, achkar12, blackburn13, blanco-canosa14}, 
Bi- \cite{comin14, da-silva-neto14, hashimoto14}, and Hg-based \cite{tabis14} compounds, 
which attracts renewed interest.  
An intriguing aspect is that the charge excitations are not closely connected with spin excitations, 
in contrast to the well-known case of La-based cuprates, where spin-charge stripes were 
discussed \cite{kivelson03}.  
Another intriguing aspect is that the charge-excitation signals develop inside 
the pseudogap state, whose origin is still a controversial issue 
in high-temperature cuprate superconductors.  

As in the hole-doped cuprates, charge excitations  
were also observed in electron-doped cuprates  Nd$_{2-x}$Ce$_{x}$CuO$_{4}$ (NCCO) 
with $x \approx 0.15$ (Ref.~\onlinecite{da-silva-neto15}). 
The situation in the electron-doped cuprates seems transparent 
in the sense that the pseudogap is not detected in a superconducting sample above its onset 
temperature ($T_c$) (Ref.~\onlinecite{misc-PGelectron}).  
Thus, the state above $T_c$ can be approximated as a normal metallic state. 
In fact, the charge excitations observed by resonant x-ray scattering (RXS) \cite{da-silva-neto15} 
are well captured by assuming a paramagnetic state in the $t$-$J$ model \cite{yamase15b}. 
Ref.~\onlinecite{yamase15b} shows that 
the observed charge excitations can be associated with a $d$-wave bond order 
and its low-energy collective excitations are realized around $\vq=(0.49 \pi,0)$ and 
$(0.84\pi, 0.84\pi)$; the former agrees with the RXS 
results \cite{da-silva-neto15} and the latter is a theoretical prediction. 
There are also individual charge excitations associated with the $d$-wave bond order, emerging 
from $\vq=(0,0)$ and $\omega=0$. 
These excitations are dominant at a relatively low energy scale of $\sim 0.1t$ 
where $t$ is a bare hopping integral in the model. 

What kind of charge excitations can then occur at a relatively high energy in the normal metallic phase? 
In particular, recent resonant inelastic x-ray scattering (RIXS) for NCCO reveals 
a dispersive signal, which has a typical energy of $0.3\, {\rm eV}$ around $\vq=(0,0)$ and  
increases to $1\, {\rm eV}$ around $\vq=(0.3\pi,0)$ \cite{ishii14,wslee14}.   
While it is not fully clear at the moment that those excitations can be 
understood also in terms of bond-order charge excitations \cite{yamase15b},  
one alternative possibility is plasmon excitations, whose energy is  typically around $1\, {\rm eV}$
in cuprates as observed in optical measurements \cite{singley01} and 
electron energy-loss spectroscopy (EELS) \cite{nuecker89,romberg90}. 
In addition, the plasmon itself is a general phenomenon in an electronic system 
and is described by the usual on-site density-density 
response function, i.e., the dielectric function. 
Hence we expect generally that plasmon excitations become relevant 
to the charge dynamics especially in a high energy region. 

In the context of cuprates, plasmons were considered as a mechanism to enhance the onset temperature 
of superconductivity in the presence of electron-phonon coupling \cite{kresin88,bill03} and 
were also studied across the metal-insulator transition \cite{vanloon14}. 
In particular, it was argued theoretically \cite{markiewicz08} that the plasmon can be detected by RIXS. 

In this paper, we study plasmon excitations in the normal metallic phase 
by including the long-range Coulomb interaction $V$, 
which is crucially important to plasmons. 
We also include interlayer hopping as a generic model for cuprate superconductors 
and study the $t$-$J$-$V$ model on a layered square-lattice. 
We invoke a large-$N$ scheme formulated in the framework of the $t$-$J$ model \cite{foussats04}, 
which successfully captures the recent RXS data for NCCO \cite{yamase15b}. 
We find that when $q_z$ (out-of-plane momentum transfer) is zero, 
the plasmon mode has a gap at $\qp=(0,0)$ and exhibits a weak 
dispersive feature away from $\qp=(0,0)$; 
here $\qp$ denotes in-plane momentum transfer.   
At finite $q_{z}$, however, the plasmon energy decreases substantially at $\qp=(0,0)$, but 
still retains a gap. This gap magnitude is scaled by 
the interlayer hopping amplitude $t_z$. 
Away from $\qp=(0,0)$ the plasmon exhibits a dispersion similar to the experimental observation 
in electron-doped cuprates \cite{wslee14}. 
Our obtained results are expected to be general for layered cuprates, implying 
a similar plasmon mode also in hole-doped cuprates, which can be tested by RIXS. 

Our theoretical study shares the importance of plasmons to the RIXS spectrum with 
Ref.~\onlinecite{markiewicz08}. The differences from the previous work \cite{markiewicz08} 
are in the inclusion of $t_z$, a functional form of the long-range Coulomb interaction, 
and a strong-coupling theory formulated in the $t$-$J$ model. 
These differences, mainly the former two, yield charge excitation spectra different 
from Ref.~\onlinecite{markiewicz08}. 

In Sec. II we describe the model and formalism. Our results are presented in Sec. III and discussed
in Sec. IV. Conclusions are given in Sec. V. 

\section{Model and Formalism}
Cuprate superconductors are layered materials where the
conducting electrons are confined to the CuO$_2$ planes, which are weakly coupled 
to each other along the $z$ direction. 
In order to address the recently observed charge excitations near $\qp=(0,0)$ \cite{ishii14, wslee14}, 
we study the $t$-$J$ model on a square lattice by including both interlayer hopping and 
the long-range Coulomb interaction, namely a layered $t$-$J$-$V$ model:  
\begin{equation}
H = -\sum_{i, j,\sigma} t_{i j}\tilde{c}^\dag_{i\sigma}\tilde{c}_{j\sigma} + 
\sum_{\langle i,j \rangle} J_{ij} \left( \vec{S}_i \cdot \vec{S}_j - \frac{1}{4} n_i n_j \right)
+\frac{1}{2} \sum_{i,j}  V_{ij} n_i n_j \, 
\label{tJV}  
\end{equation}
where the sites $i$ and $j$ run over a three-dimensional lattice. 
The hopping $t_{i j}$ takes a value $t$ $(t')$ between the first (second) nearest-neighbors 
sites on the square lattice whereas 
hopping integrals between the layers are scaled by $t_z$ and 
we will specify the out-of-plane dispersion later [see \eq{Eperp}].  
$\langle i,j \rangle$ indicates a nearest-neighbor pair of sites on the square lattice, and 
we consider the exchange interaction only inside the plane, namely $J_{i j}=J$ 
because the exchange term between adjacent planes is much smaller than $J$
(Ref.~\onlinecite{thio88}).
$V_{ij}$ is the long-range Coulomb interaction on the lattice and is given in momentum space by 
\be
V(\vq)=\frac{V_c}{A(q_x,q_y) - \cos q_z} \,,
\label{LRC}
\ee
where $V_c= e^2 d(2 \epsilon_{\perp} a^2)^{-1}$ and 
\be
A(q_x,q_y)=\frac{\tilde{\epsilon}}{(a/d)^2} (2 - \cos q_x - \cos q_y)+1 \,.
\ee
These expressions are easily obtained by solving Poisson's equation on a lattice \cite{becca96}.  
Here $\tilde{\epsilon}=\epsilon_\parallel/\epsilon_\perp$,  
and $\epsilon_\parallel$ and $\epsilon_\perp$ are the 
dielectric constants parallel and perpendicular to the planes, respectively, and they are positive;  
$V(\vq)$ is thus positive for any $\vq$; 
$a$ and $d$ are the lattice spacing in the planes and between the 
planes, respectively; $e$ is the electric charge of electrons.
In the present study, momentum is measured in units of the inverse of the lattice constants
in each direction, namely $a^{-1}$ along the $x$ and $y$ direction and $d^{-1}$ along the $z$ direction. 
$\tilde{c}^\dag_{i\sigma}$ ($\tilde{c}_{i\sigma}$) is 
the creation (annihilation) operator of electrons 
with spin $\sigma$ ($\sigma = \downarrow$,$\uparrow$) 
in the Fock space without double occupancy. 
$n_i=\sum_{\sigma} \tilde{c}^\dag_{i\sigma}\tilde{c}_{i\sigma}$ 
is the electron density operator and $\vec{S}_i$ is the spin operator.

We analyze the model (\ref{tJV}) in terms of a large-$N$ expansion formulated in Ref.~\onlinecite{foussats04}
for Hubbard operators.
While it was formulated for the purely two-dimensional $t$-$J$ model, it is straightforward to 
extend it to the present layered model. 
Since the method was described in previous papers \cite{foussats04,bejas12,bejas14}, here we reproduce only the main 
details.
In the large-$N$ scheme, the charge excitations are described by 
a six-component bosonic field, namely $\delta X^{a}$ with $a=1,2,\cdots,6$. $\delta X^{1}$ describes 
on-site charge fluctuations and $\delta X^{2}$ fluctuations around the 
mean value of a Lagrange multiplier associated with the constraint of non-double occupancy. 
$\delta X^{3}$ and $\delta X^{4}$ ($\delta X^{5}$ and $\delta X^{6}$) describe 
real (imaginary) parts of bond-field fluctuations along the $x$ and $y$ directions, respectively; 
the expectation value of the bond field is specified later [\eq{Delta}]. The six-component boson 
propagator is then given by a $6 \times 6$ matrix. 
At leading order, the inverse of the propagator is given by  
\be
D^{-1}_{ab}(\vq,\mathrm{i}\omega_n)
= [D^{(0)}_{ab}(\vq,\mathrm{i}\omega_n)]^{-1} - \Pi_{ab}(\vq,\mathrm{i}\omega_n)\,.
\label{dyson}
\ee
Here $a$ and $b$ run from 1 to 6, $\omega_{n}$ is a bosonic Matsubara frequency, and 
$D^{(0)}_{ab}(\vq,\mathrm{i}\omega_n)$ is a bare bosonic propagator
\begin{widetext}
\begin{equation} \label{D0inverse}
[D^{(0)}_{ab}(\vq,\mathrm{i}\omega_n)]^{-1} = N 
\left(
\begin{array}{llllll}
\frac{\delta^2}{2} \left( V(\vq)-J(\vq)\right) 
& \frac{\delta}{2} & 0 & 0 & 0 & 0\\
\frac{\delta}{2} & 0 & 0 & 0 & 0 & 0\\
0 & 0 & \frac{4\Delta^2}{J} & 0 & 0 & 0\\
0 & 0 & 0 & \frac{4\Delta^2}{J} & 0 & 0\\
0 & 0 & 0 & 0 & \frac{4\Delta^2}{J} & 0\\
0 & 0 & 0 & 0 & 0 & \frac{4\Delta^2}{J}
\end{array}
\right) \; ,
\end{equation}
\end{widetext}
with $J(\vq) = \frac{J}{2} (\cos q_x +  \cos q_y)$.   
$\Pi_{ab}(\vq,\mathrm{i}\omega_n)$ are the bosonic self-energies at leading order:  
\begin{eqnarray}
&& \Pi_{ab}(\vq,\mathrm{i}\omega_n)
            = -\frac{N}{N_s N_z}\sum_{\vk} h_a(\vk,\vq,\varepsilon_\vk-\varepsilon_{\vk-\vq}) 
            \frac{n_F(\varepsilon_{\vk-\vq})-n_F(\varepsilon_\vk)}
                                  {\mathrm{i}\omega_n-\varepsilon_\vk+\varepsilon_{\vk-\vq}} 
            h_b(\vk,\vq,\varepsilon_\vk-\varepsilon_{\vk-\vq}) \nonumber \\
&& \hspace{25mm} - \delta_{a\,1} \delta_{b\,1} \frac{N}{N_s N_z}
                                       \sum_\vk \frac{\varepsilon_\vk-\varepsilon_{\vk-\vq}}{2}n_F(\varepsilon_\vk) \; .
                                       \label{Pi}
\end{eqnarray}
The factor $N$ in front of Eqs.~(\ref{D0inverse}) and (\ref{Pi})  comes from 
the sum over the $N$ fermionic channels after the extension of the spin index 
$\sigma$ from $2$ to $N$, whereas $N_s$ and $N_z$ are the total 
number of lattice sites on the square lattice and the number of layers along the $z$ direction, respectively. 
The electronic dispersion $\varepsilon_{\vk}$ may be written as 
\be
\varepsilon_{\vk} = \varepsilon_{\vk}^{\parallel}  + \varepsilon_{\vk}^{\perp} 
\label{Ek}
\ee
where the in-plane dispersion $\varepsilon_{\vk}^{\parallel}$ and the out-of-plane dispersion 
$\varepsilon_{\vk}^{\perp}$ are given by, respectively,  
\bea
&&\varepsilon_{\vk}^{\parallel} = -2 \left( t \frac{\delta}{2}+\Delta \right) (\cos k_{x}+\cos k_{y})-
4t' \frac{\delta}{2} \cos k_{x} \cos k_{y} - \mu \,,\\
\label{Epara}
&&\varepsilon_{\vk}^{\perp} = 2 t_{z} \frac{\delta}{2} (\cos k_x-\cos k_y)^2 \cos k_{z}  \,.
\label{Eperp}
\eea
The functional form $ (\cos k_x-\cos k_y)^2$ in $\varepsilon_{\vk}^{\perp}$ is frequently invoked for cuprates \cite{andersen95}. 
Here $\delta$ is doping rate, $\mu$ the chemical potential, and $\Delta$ the mean-field 
value of the bond-field. For a given $\delta$, $\mu$ and $\Delta$ are determined self-consistently by solving 
\bea 
&&\Delta = \frac{J}{4N_s N_z} \sum_{\vk} (\cos k_x +\cos k_y ) n_F(\varepsilon_\vk)\,, 
\label{Delta} \\
&&(1-\delta)=\frac{2}{N_s N_z} \sum_{\vk} n_F(\varepsilon_\vk)\,, 
\eea
where $n_F$ is the Fermi function.  The six-component vertex $h_a$ in \eq{Pi} is given by 
\begin{widetext}
\begin{align}
 h_a(\vk,\vq,\nu) =& \left\{
                   \frac{2\varepsilon_{\vk-\vq}+\nu+2\mu}{2}+
                   2\Delta \left[ \cos\left(k_x-\frac{q_x}{2}\right)\cos\left(\frac{q_x}{2}\right) +
                                  \cos\left(k_y-\frac{q_y}{2}\right)\cos\left(\frac{q_y}{2}\right) \right];1;
                 \right. \nonumber \\
               & \left. -2\Delta \cos\left(k_x-\frac{q_x}{2}\right); -2\Delta \cos\left(k_y-\frac{q_y}{2}\right);
                         2\Delta \sin\left(k_x-\frac{q_x}{2}\right);  2\Delta \sin\left(k_y-\frac{q_y}{2}\right)
                 \right\} \; . 
\end{align}
\end{widetext}

We compute the on-site density-density correlation function in the present 
large-$N$ framework. After summing all contributions up to $O(1/N)$, we obtain 
\begin{eqnarray}\label{chi}
\chi^c({\bf q},{\rm i}\omega_n)=
-N {\left( \frac{\delta}{2} \right)}^2 D_{11}({\bf q},{\rm i}\omega_n)\,.
\end{eqnarray}
Thus, the density-density correlation function is connected with the component $(1,1)$ of the $D_{ab}$.

Although the physical value is $N=2$, the large-$N$ expansion has several 
advantages over usual perturbations theories. 
First, in contrast to usual random-phase approximation (RPA) which is valid 
in the weak coupling regime, $\chi^c$ is not obtained as perturbation of any physical
parameters of the model (\ref{tJV}). 
This difference is crucial near half-filling where strong correlations are expected to be important. 
In fact,  charge degrees of freedom are generated by carrier doping into a Mott insulator and thus 
the long-range part of the Coulomb interaction should vanish at half-filling. 
This feature is captured already in the present leading order theory 
[see the $(1,1)$ component in \eq{D0inverse}], but not in weak coupling \cite{hoang02}. 
Second, it was shown \cite{merino03} that the large-$N$
expansion reproduces well charge excitations obtained by exact diagonalization. 
Moreover, the present formalism nicely captures \cite{yamase15b} the charge order 
recently found by RXS \cite{da-silva-neto15}. 
We therefore believe that the large-$N$ theory is a powerful approach to explore 
charge excitations in cuprates.

\section{Results}

In what follows we present results for the parameters $J/t=0.3$ and $t'/t=0.30$ 
which are appropriate for electron-doped cuprates \cite{yamase15b}. 
The number of layers we take are 30, which is sufficiently large. 
We take $t_z = 0.1 t$, for which we checked that the topology of the Fermi surface 
is the same as that for $t_z=0$ in the doping region we will consider. 
Since that choice of $t_z$ may be rather arbitrary, 
we will present the $t_z$ dependence of our main results. 
Concerning the long-range Coulomb interaction [\eq{LRC}], we choose 
$d/a=1.5$ (Ref.~ \onlinecite{misc-d}) with $a=4$~{\AA}; while the choice of $\epsilon_{\parallel}$ and 
$\epsilon_{\perp}$ is not universal among theoretical papers \cite{becca96,prelovsek99}, 
we respect the experimental data \cite{timusk89} and choose $\epsilon_\parallel=4 \epsilon_0$ and 
$\epsilon_\perp=2 \epsilon_0$ with $\epsilon_0$ being the dielectric constant in vacuum. 
We present results mainly for $\delta=0.15$, which allows us a direct comparison with recent experiments in the 
electron-doped cuprates \cite{ishii14,wslee14}.

We compute the spectral weight of the density-density correlation function Im$\chi^{c}(\vq,\omega)$ after 
analytical continuation 
\be
{\rm i} \omega_n \rightarrow \omega + {\rm i} \Gamma
\label{continuation}
\ee
in \eq{chi}. The value of $\Gamma$ is positive and is in principle infinitesimally small. 
We choose $\Gamma = 10^{-4}t$ for numerical convenience in most of cases 
and will clarify the $\Gamma$ dependence 
of our main results. 

\begin{figure}
\centering
\includegraphics[width=16cm]{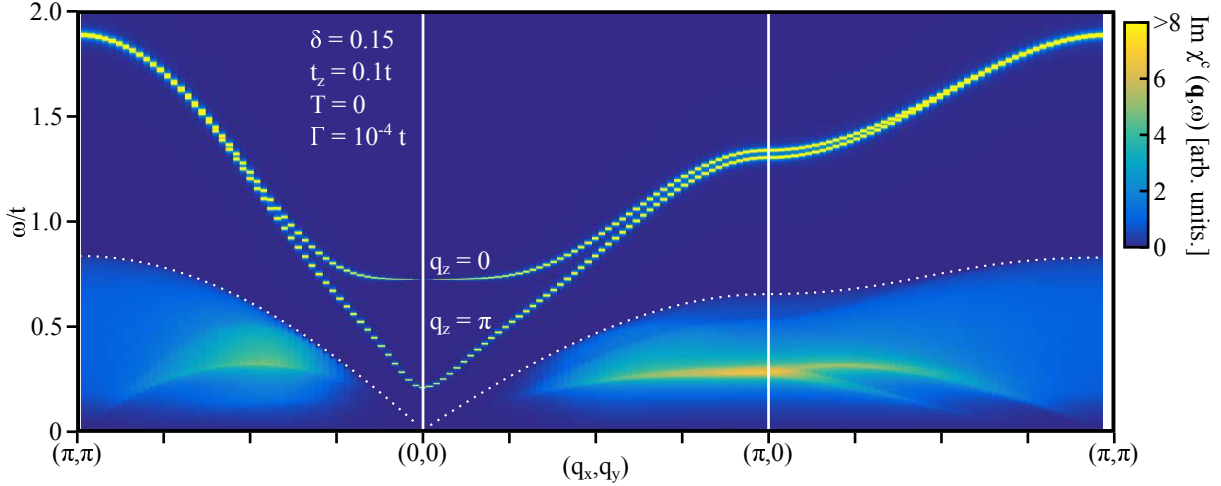}
\caption{(Color online) Spectral weight of the density-density correlation Im$\chi^{c}(\vq,\omega)$ in the 
plane of energy $\omega$ and in-plane momentum $\qp$ along
$(\pi,\pi)-(0,0)-(\pi,0)-(\pi,\pi)$ for $q_z=0$ and $\pi$. 
The dotted line denotes the upper boundary of a particle-hole continuum for $q_z=0$. 
The spectral intensity of the plasmon is by far higher than the continuum and is cut 
at $8$ to get a better contrast for the continuum. 
}
\label{map}
\end{figure}
Figure~\ref{map} shows a map of the spectral weight Im$\chi^{c}(\vq,\omega)$ in the plane of 
excitation energy $\omega$ and in-plane momentum $\qp$ along 
symmetry axes of $(\pi,\pi)-(0,0)-(\pi,0)-(\pi,\pi)$. 
Below $\omega \sim 0.8t$, there is a particle-hole continuum coming from 
individual charge excitations. 
The continuum does not depend much on $q_z$ except that the spectral weight 
is slightly enhanced along $(\pi,\pi)-(0,0)-(\pi,0)$ for $q_z=\pi$ compared with that for $q_z=0$. 
In the figure the continuum for $q_z=0$ is presented. 
We find no strong spectral weight near zero energy, implying that there is no charge order 
tendency associated with $\chi^{c}(\vq,0)$, namely 
the $(1,1)$ component in the bosonic propagator of \eq{dyson}.  
Instead various charge orders can occur in the components specified by $a=b=3, 4, 5$, and $6$ 
as shown previously \cite{bejas12,bejas14}. 
In a high energy region, there is a sharp and strong weight for $q_z=0$. This is a particle-hole 
bound state realized above the continuum, namely a plasmon.
The plasmon energy at $\qp=(0,0)$ is around $\omega_{p}=0.7t$. 
Its dispersion is characterized by 
\be
\omega(\qp, q_z=0)=\omega_{p} + a_2 q_{\parallel}^{2} + \cdots \, .
\label{plasmon}
\ee
The dispersion looks quite flat near $\qp=(0,0)$, suggesting a very small value of $a_2$. 
The coefficient of $a_2$ is approximated well by a formula obtained in the electron gas model \cite{mahan}, 
which predicts $a_2= (3/10) v_{F}^{2}/\omega_p$ with $v_{F}$ being the Fermi velocity; 
we may consider an average of the Fermi velocity in our case. 
Because the bare hopping integrals $t$, $t'$, and $t_z$ are renormalized by a factor of $\delta/2$ in \eq{Ek}, 
our Fermi velocity becomes rather small at $\delta=0.15$. 
This is a major reason why the plasmon dispersion 
is very flat near $\qp=(0,0)$ in \fig{map}.  

The plasmon dispersion changes drastically around $\qp=(0,0)$ when $q_{z}$ becomes finite. 
As a representative, we plot the plasmon dispersion for $q_{z}=\pi$ in \fig{map}. 
While the plasmon dispersion remains essentially the same as that for $q_z=0$ far away from $\qp=(0,0)$, 
the plasmon energy softens substantially near $\qp=(0,0)$ and exhibits a strong dispersion 
there, in sharp contrast to that for $q_z=0$. 
The plasmon has a quadratic dispersion near $\qp=(0,0)$.

\begin{figure}[t]
\centering
\includegraphics[width=7cm]{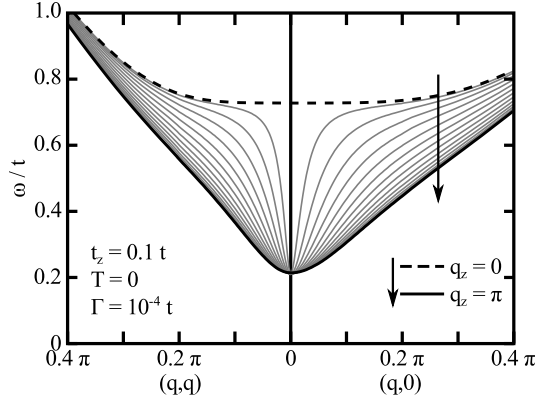}
\caption{$q_z$ dependence of the plasmon dispersion around $\qp=(0,0)$ 
along $(0.4\pi,0.4\pi)$-$(0,0)$-$(0.4\pi,0)$. 
$q_z$ is changed from zero to $\pi$ with a step of $\pi/15$.  
This value of $\pi/15$ comes from our resolution of $q_z$ 
since we take 30 layers in our model.  
}
\label{qz-depend}
\end{figure}

Figure~\ref{qz-depend} shows how the plasmon dispersion changes as increasing $q_z$ 
from zero to $\pi$. Recalling that plasmons originate from the singularity of the long-range 
Coulomb interaction with $\sim 1/(\frac{\tilde{\epsilon}}{(a/d)^2}  \qp^2+q_z^2)$ 
[see \eq{LRC}] at small momenta \cite{mahan}, 
the plasmon energy at $\qp=(0,0)$ changes discontinuously once $q_z$ becomes finite 
in a layered system.  Because of this singularity, 
the plasmon dispersion becomes very sensitive to a slight change of 
$q_z$ from zero for a small $\qp$. 
This strong $q_z$ dependence of the plasmon mode 
was also obtained in literature \cite{kresin88,bill03,markiewicz08}, although $t_z=0$ was assumed there. 

\begin{figure}[t]
\centering
\includegraphics[width=7cm]{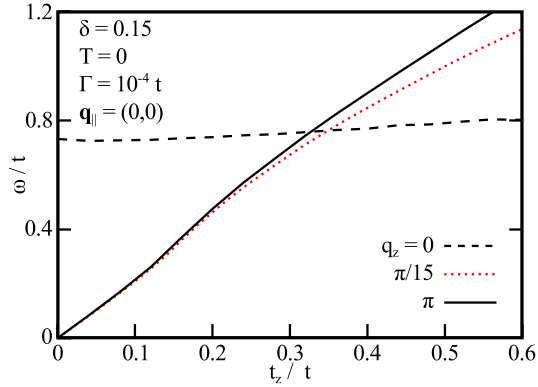}
\caption{(Color online) $t_z$ dependence of the plasmon energy at $\qp=(0,0)$
for $q_z=0, \pi/15$, and $\pi$. 
}
\label{tz-depend}
\end{figure}

In \fig{tz-depend} we shows the $t_z$ dependence of the 
plasmon energy at $\qp=(0,0)$ for several choices of $q_z$. 
At $q_z=0$, the plasmon energy is almost independent of $t_z$. 
For a finite $q_z$, its $t_z$ dependence becomes completely different from 
that for $q_z=0$. As expected from \fig{qz-depend},  
the plasmon energy at $\qp=(0,0)$ drops discontinuously for a small $t_z$ 
and exhibits almost the same $t_z$ dependence for both $q_z=\pi/15$ and $\pi$. 
The plasmon energy vanishes at $t_z=0$, in agreement with literature \cite{kresin88,bill03,markiewicz08}. 
As increasing $t_z$, the plasmon energy increases monotonically, with a linear dependence for a small $t_z$.  
Interestingly the plasmon energy becomes higher 
than that at $q_z=0$ for $t_z > 0.30t$ (see Appendix~A). 
In Figs.~\ref{map} and \ref{qz-depend} 
we take $t_z=0.1t$ and thus the plasmon energy has $\omega \approx 0.2t$ at $\qp=(0,0)$  
for $q_z=\pi$ and $0.7t$ for $q_z=0$. 

\begin{figure}
\centering
\includegraphics[width=7cm]{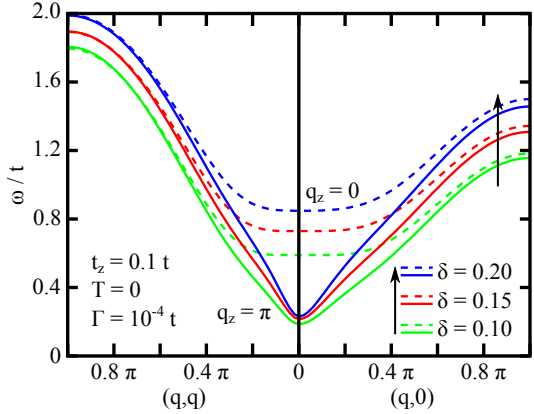}
\caption{(Color online) Plasmon dispersions for $q_z=0$ (dashed line) and $\pi$ (solid line) for several
choices of doping rate $\delta$ along $(\pi,\pi)$-$(0,0)$-$(\pi,0)$. 
}
\label{dispersion}
\end{figure}
The plasmon dispersions for $q_z=0$ and $\pi$ are summarized in \fig{dispersion} by performing 
calculations at different doping rates. The plasmon energy tends to increase with increasing 
doping. 
Since the plasmon energy at $\vq=(0,0,0)$ is given by $\omega_{p}^{2}=\frac{n_0 e^2}{\epsilon_{0} m}$ 
in the electron gas model \cite{mahan}, where $n_0$ is the electron density and 
$m$ is an electron mass, 
the hardening of the plasmon with carrier doping might be surprising since we would assume $n_0 = 1-\delta$. 
However, electronic charge is introduced by carrier doping in the  $t$-$J$ model and there is no 
plasmon at half-filling \cite{prelovsek99}. 
We thus expect that the plasmon energy becomes higher with carrier doping at least 
close to half filling as shown in \fig{dispersion}. See also Appendix~B, where the plasmon energy 
is studied in the entire doping region $0 < \delta < 1$.

\begin{figure}[ht]
\centering
\includegraphics[width=7cm]{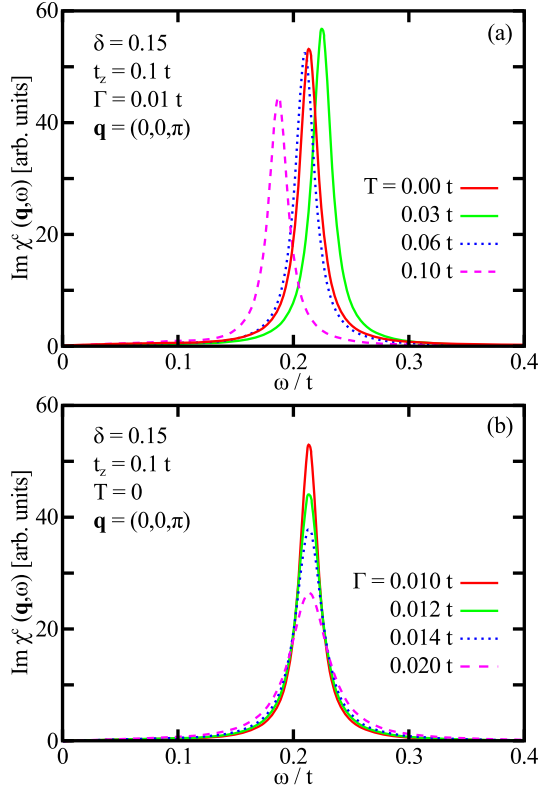}
\caption{(Color online) (a) Temperature dependence of Im$\chi^{c}(\vq,\omega)$ at $\vq=(0,0,\pi)$  for $\Gamma=0.01t$. 
(b) Im$\chi^{c}(\vq,\omega)$ at $T=0$ for several choices of $\Gamma$. 
}
\label{T-G-depend}
\end{figure}
So far we have focused on calculations at temperature $T=0$. 
Now we study the effect of temperature. 
In \fig{T-G-depend} (a) we plot Im$\chi^{c}(\vq,\omega)$ at $\vq=(0,0,\pi)$ for several choices of 
temperatures. Surprisingly, not only the peak width but also the peak height is 
essentially unchanged with increasing $T (< 0.06 t)$, although the peak position shifts slightly 
in a non-monotonic way: with raising temperature, the peak energy first increases a little bit  
and then decreases above $T \approx 0.025 t$. 
This behavior reflects a weak temperature dependence of Re$\chi^{c}(\vq,\omega)$ around $\omega \sim 0.2t$.  
If we invoke an unphysically large $T$, e.g., more than 500 K, which may correspond 
to $T > 0.05 t$ if $t/2=500$ meV \cite{hybertsen90,misc-factor2}, 
the plasmon energy decreases and its peak height is gradually 
suppressed with further increasing $T$, keeping the peak width almost unchanged. 

On the other hand, the peak height is quite sensitive to the damping factor $\Gamma$ 
introduced in \eq{continuation}. Figure~\ref{T-G-depend} (b) shows Im$\chi^{c}(\vq,\omega)$  
at $\vq=(0,0,\pi)$ at $T=0$ for several choices of $\Gamma$. 
With increasing  $\Gamma$ the peak loses intensity and becomes slightly broader.

While we have taken \eq{Eperp} as the out-of-plane dispersion, 
we made the same calculations also for 
\be
\varepsilon_{\vk}^{\perp} = 2 t_{z} \frac{\delta}{2} \cos k_{z}  
\label{Eperp2}
\ee
and found qualitatively the same results. We thus expect that our major results 
do not depend on details of the out-of-plane dispersion. 

\section{Discussions}
We have studied plasmon modes, which are particle-hole bound states realized above the continuum 
(see \fig{map}), in a layered $t$-$J$ model by including the long-range Coulomb interaction. 
The plasmon dispersion is the usual optical mode at $q_z=0$, but the dispersion changes 
drastically by shifting $q_z$ slightly from zero (\fig{qz-depend}). 
The resulting dispersion becomes strong 
near $\qp=(0,0)$ and its energy at $\qp=(0,0)$ is scaled by $t_z$ (\fig{tz-depend}).  

Since the value of $q_z$ is usually finite and well away from zero in RIXS, 
we consider that the mode observed by Lee {\it et al.} \cite{wslee14} can be associated 
with the plasmon mode with a finite $q_z$, typically with $q_z=\pi$ (Figs.~\ref{map} , \ref{qz-depend}, 
and \ref{dispersion}). 
The gap magnitude at $\qp=(0,0)$ obtained by the experiment \cite{wslee14} 
is therefore interpreted  to be proportional to the interlayer coupling $t_z$ (\fig{tz-depend}).  
From a viewpoint of this plasmon scenario, 
it is interesting to clarify to what extent the present theory captures quantitative aspects. 
Our result suggests that $\omega \approx 2 t_{z}$ for a small $t_z$ 
in \fig{tz-depend} and thus 
$t_{z}^{\rm exp} \approx \frac{0.3\, {\rm eV}}{2} = 0.15 \, {\rm eV}$ 
as a bare hopping amplitude. 
Since the actual band width is renormalized by a factor of $\delta/2$ [see \eq{Eperp}], 
the obtained out-of-plane hopping at $(\pi,0)$ becomes 90 meV. 
This value is a factor of three larger than the 30 meV obtained for single-layer  
cuprates (without apical oxygen atoms of Cu) by LDA \cite{andersen95}. 
On the other hand, the slope of our plasmon 
dispersion near $\qp=(0,0)$ for a finite $q_z$ (Figs.~\ref{map}, \ref{qz-depend}, and \ref{dispersion} for $q_z=\pi$) 
agrees with experimental observation within a factor of 1.5. 
In optical measurements \cite{singley01} and EELS \cite{nuecker89, romberg90}, 
the value of $q_z$ is zero. 
We thus expect naturally a large plasmon energy of the order of $t$. 
The $q^{2}$ dependence of the plasmon \cite{nuecker89,romberg90} is nicely captured, 
but our coefficient $a_2$ [see \eq{plasmon}] is smaller than the experiments. 
In spite of these quantitative differences between the present theory and the experiments,  
our interpretation has the advantage to provide a consistent understanding of the mode 
recently found by Lee {\it et al.} \cite{wslee14} and the mode observed by optical measurements \cite{singley01} 
and EELS \cite{nuecker89,romberg90} by invoking a different value of $q_z$. 
This possibility was not considered by Lee {\it et al.} \cite{wslee14}. 

The mode observed by Lee {\it et al.} \cite{wslee14} exhibits a hardening with increasing carrier doping 
away from $\qp=(0,0)$. 
This hardening is captured clearly in a large $|\qp|$ region for $q_z=\pi$ as shown in \fig{dispersion}. 
In the vicinity of $\qp=(0,0)$, on the other hand, Lee {\it et al.} \cite{wslee14} reported a softening 
by comparing data at $x=0.147$ and $0.166$, which, however, looks quite subtle,  
whereas we have found a very small change even between $\delta=0.15$ and $0.20$.  
More comprehensive measurements for various dopings near $\qp=(0,0)$ may be useful 
to make a further comparison. 
For $q_z=0$, on the other hand, the hardening of the plasmon with increasing $\delta$ 
is consistent with experimental observation \cite{hwang07}. 

The temperature dependence of the charge excitation intensity obtained by Lee {\it et al.} \cite{wslee14} could be 
understood within our framework. As shown in \fig{T-G-depend}, 
the peak intensity of the plasmon is rather delicate. Within a realistic range 
of temperature in \fig{T-G-depend} (a) ($< 0.05 t$), 
not only the peak position but also the peak height and peak width do not change much. 
However, the peak height strongly depends on the damping factor $\Gamma$ 
as shown in \fig{T-G-depend}~(b). While $\Gamma$ is introduced in the  analytical continuation 
in \eq{continuation}, $\Gamma$ may mimic a self-energy effect, especially for 
a relatively high energy region that we are interested in, because 
the imaginary part of the self-energy may not depend much on momentum and energy there. 
This feature allows us to consider the first term of the self-energy, namely a constant term, 
in Taylor series with respect to momentum and energy around the plasmon energy 
at $\vq=(0,0,\pi)$. 
If we invoke that $\Gamma$ decreases with decreasing temperature, we may understand 
the temperature dependence of the charge excitation intensity at $\delta =0.166$ observed 
by Lee {\it et al.} \cite{wslee14}. In contrast, the corresponding data 
at $\delta=0.147$ in the experiment \cite{wslee14} exhibits a much weaker temperature dependence,  
which can be due to a smaller temperature dependence of $\Gamma$ in the corresponding temperature 
region.

In Ref.~\onlinecite{ishii14}, a charge excitation peak near $\qp=(0,0)$, which seems 
the same signal reported in Ref.~\onlinecite{wslee14}, 
was interpreted in terms of individual particle-hole excitations, not a collective mode such as 
the plasmon. If the individual excitations are responsible for that, one would expect 
a gapless excitation spectrum at $\qp=(0,0)$. 
It is therefore worth performing RIXS measurements more in detail near $\qp=(0,0)$ 
to clarify whether the gap is indeed present at $\qp=(0,0)$ or not, which is a crucial test of 
our interpretation of plasmons. 

As frequently seen in standard textbooks in condensed matter physics \cite{mahan}, 
a plasmon mode is often presumed to overdamp 
as momentum transfer becomes  large because of the mixture with the 
particle-hole continuum. 
Such a result is obtained in an electron gas in a continuum model. 
On the other hand, in a layered system defined on a lattice, which 
is more realistic to cuprates, we find that the plasmon 
is well separated from the particle-hole continuum up to 
the zone boundary (see \fig{map}). 
It is, however, noted that the present one-band model contains only intraband scattering processes 
and in general there should be also various interband scattering processes in real materials. 
The plasmon is thus likely overdamped due to the mixture of the interband excitations. 
At present, available data \cite{ishii14}  suggest that 
the interband spectral weight is dominant only rather close to 
$\qp=(0,0)$ around $\omega = 2\, {\rm eV}$. 
Hence it is interesting to test how the mode observed by Lee {\it et al.} \cite{wslee14} 
disperses for a large $|\qp|$.

The two experimental works, Refs.~\onlinecite{ishii14} and \onlinecite{wslee14},  
should not be mixed with Ref.~\onlinecite{da-silva-neto15}. 
For the former studies (Refs.~\onlinecite{ishii14} and \onlinecite{wslee14}), 
we here propose plasmon modes  
to interpret the charge-excitation peak observed around $\qp=(0,0)$ in a relatively high energy region. 
For the latter (Ref.~\onlinecite{da-silva-neto15}), 
the charge order signal was observed at $\qp \approx (0.48\pi,0)$, far away from 
$\qp=(0,0)$. This data can be interpreted in terms of 
$d$-wave bond-order charge excitations described by the third and fourth components of $D_{ab}$ 
in \eq{dyson}, as shown in our previous work \cite{yamase15b}. 
While the long-range Coulomb interaction was not considered previously \cite{yamase15b},  
its impact on the $d$-wave bond-order is expected to be very weak since the $D_{11}$ and  
$D_{ab}$ with $a,b=3,4$ are almost decoupled with each other\cite{bejas12,bejas14}.

The present study can be highlighted also through a comparison with previous theoretical 
studies of plasmons in cuprates. 
First, we share a strong $q_z$ dependence of the plasmon dispersion with 
early theoretical studies \cite{kresin88,bill03,markiewicz08}. 
Second, previous calculations \cite{kresin88,bill03,markiewicz08,vanloon14} 
did not take the interlayer hopping $t_z$ into account 
and reported a gapless dispersion for a finite $q_z$. 
However, we have shown a striking insight that the plasmon mode has a sizable gap 
with a quadratic dispersion near $\qp=(0,0)$ for a finite $q_z$ 
because of a finite interlayer hopping $t_z$. 
This aspect is very important to study charge excitations in cuprates. 
Third, Ref.~\onlinecite{markiewicz08} predicts that the plasmon for $q_z=0$ has a very steep dispersion 
around $\qp=(0,0)$ 
whereas we have obtained a very flat dispersion (Figs.~\ref{map} and \ref{dispersion} for $q_z=0$). 
Fourth, previous studies \cite{kresin88,bill03,markiewicz08} are based on 
a weak coupling theory and consider the presence of the 
long-range Coulomb interaction even at half-filling, whereas, consistent with a phenomenology of 
a Mott insulator \cite{vanloon14},  
the long-range Coulomb interaction vanishes at half-filling in the present large-$N$ theory. 

Our obtained results (\fig{map}) are expected 
to be a representative of cuprate superconductors in general. 
Although a mode similar to that observed by Lee {\it et al.} \cite{wslee14} has not been reported yet 
in hole-doped cuprates, we expect that a similar mode should be present 
and can be detected by RIXS at least in a normal state in a heavily doped region 
where the pseudogap effect is substantially weakened. 
A recent RIXS study by Ishii {\it et al.} \cite{misc-ishii} suggests the presence of charge excitations 
near $\qp=(0,0)$ in La$_{2-x}$Sr$_{x}$CuO$_{4}$ similar to that observed in NCCO \cite{ishii14,wslee14}.

\section{Conclusions}
In summary, we have studied plasmon excitations in the $t$-$J$ model by including 
an interlayer coupling and the long-range Coulomb interaction as a generic model 
of the layered cuprate superconductors. 
We have found that 
the plasmon dispersion exhibits a strong $q_{z}$ dependence in the vicinity of $\qp=(0,0)$ 
(Figs.~\ref{map} and \ref{qz-depend}). 
For $q_z=0$ the plasmon has a large energy gap of the order of $t$ near $\qp=(0,0)$, 
which may be associated with the mode observed in optics\cite{singley01} and 
EELS \cite{nuecker89,romberg90}. 
For a finite $q_z$, on the other hand, the plasmon energy is substantially suppressed to a 
low energy at $\qp=(0,0)$ and exhibits a strong dispersive feature 
(see Figs.~\ref{map}, \ref{qz-depend}, and \ref{dispersion}). 
We have also found that the plasmon energy at $\qp=(0,0)$ for a finite $q_z$ 
is scaled by the interlayer hopping amplitude $t_z$ (\fig{tz-depend}). 
Since $q_z$ is in general finite in RIXS measurements, the mode observed by Lee {\it et al.} \cite{wslee14} 
can be associated with the plasmon mode with a finite $q_z$. 
The present results should not be mixed with a recent observation of a short-range charge order by 
RIXS \cite{da-silva-neto15}. 
Within essentially the same large-$N$ framework as the present one, 
that charge-order peak was interpreted in terms of the $d$-wave bond-order 
correlation function \cite{yamase15b},  
which has a dominant spectral weight far away from $\qp=(0,0)$ and in energy much lower than the plasmon. 
While we have presented results for a parameter set 
appropriate for electron-doped cuprates, our results are expected to be applicable to hole-doped cuprates 
at least in the normal state in the overdoped side where the effect of the pseudogap is substantially weakened, 
implying that the plasmon mode will be observed by RIXS also in hole-doped cuprates. 

\acknowledgments
The authors thank K. Ishii for very fruitful discussions about RIXS measurements in cuprates. 
H.Y. acknowledges support by a Grant-in-Aid for Scientific Research Grant No.
JP15K05189.


\appendix
\section{Plasmons for a large {$\boldsymbol t_z$}} 
\begin{figure}[ht]
\centering
\includegraphics[width=7cm]{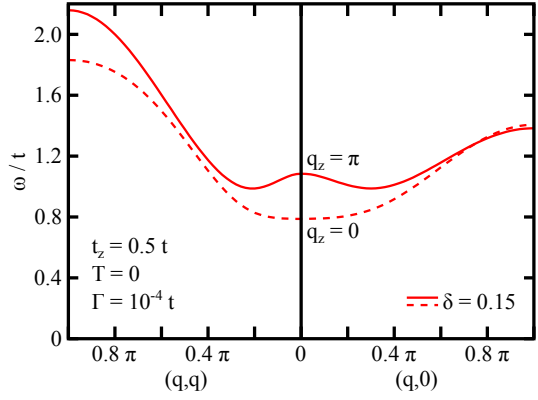}
\caption{(Color online) Plasmon dispersions for $q_z=0$ (dashed line) and $\pi$ (solid line)
for $\delta=0.15$ and $t_z=0.5t$ along $(\pi,\pi)$-$(0,0)$-$(\pi,0)$. 
}
\label{tz-depend2}
\end{figure}
While a realistic value of $t_z$ in cuprates is expected much smaller than $t$ in \fig{tz-depend}, 
it is interesting that the plasmon energy for $q_z=\pi$ becomes larger than that for $q_z=0$ 
for a large $t_z$. This is demonstrated in \fig{tz-depend2} by computing 
plasmon dispersions for $t_z=0.5t$ as a function of $\qp$, which   
may be applicable to a layered material with strong three dimensionality. 
Interestingly, the dispersion for $q_z=\pi$ has a negative curvature at $\qp=(0,0)$ 
and forms local minima at $\qp \approx (0.3\pi,0)$ and $(0.2\pi, 0.2\pi)$. 
For a larger $|\qp|$, the dispersion for $q_z=\pi$ shows a dependence 
similar to that for $q_z=0$. 

\section{Doping dependence of plasmon energy}
\begin{figure}[ht]
\centering
\includegraphics[width=7cm]{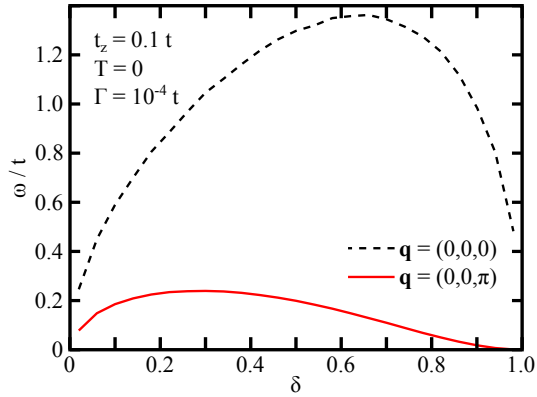}
\caption{(Color online) Plasmon energy at $\vq=(0,0,0)$ (dashed line) and $(0,0,\pi)$ (solid line) 
as a function of doping. 
} 
\label{delta-depend}
\end{figure}

In \fig{dispersion}, we have presented the plasmon dispersions for several choices 
of $\delta$. Figure~\ref{delta-depend} is a complementary result by plotting 
the plasmon energy at $\qp=(0,0)$ as a function of $\delta$. 
The plasmon energy for both $q_z=0$ and $\pi$ increases with $\delta$ at least 
in a doping region realistic to cuprates ($\delta \lesssim  0.30$). 
This is consistent with a previous work in the $t$-$J$ model \cite{prelovsek99}, 
where the plasmon energy for $q_z=0$ was studied as a  function of $\delta$. 
The plasmon energy forms a maximum and then decreases for a 
high $\delta$. This is easily understood by noting that the electronic band 
becomes empty at $\delta=1$, where the plasmon energy would vanish.


\begin{thebibliography}{39}
\expandafter\ifx\csname natexlab\endcsname\relax\def\natexlab#1{#1}\fi
\expandafter\ifx\csname bibnamefont\endcsname\relax
  \def\bibnamefont#1{#1}\fi
\expandafter\ifx\csname bibfnamefont\endcsname\relax
  \def\bibfnamefont#1{#1}\fi
\expandafter\ifx\csname citenamefont\endcsname\relax
  \def\citenamefont#1{#1}\fi
\expandafter\ifx\csname url\endcsname\relax
  \def\url#1{\texttt{#1}}\fi
\expandafter\ifx\csname urlprefix\endcsname\relax\def\urlprefix{URL }\fi
\providecommand{\bibinfo}[2]{#2}
\providecommand{\eprint}[2][]{\url{#2}}

\bibitem[{\citenamefont{Ghiringhelli et~al.}(2012)\citenamefont{Ghiringhelli,
  Le~Tacon, Minola, Blanco-Canosa, Mazzoli, Brookes, Luca, Frano, Hawthorn, He
  et~al.}}]{ghiringhelli12}
\bibinfo{author}{\bibfnamefont{G.}~\bibnamefont{Ghiringhelli}},
  \bibinfo{author}{\bibfnamefont{M.} \bibnamefont{Le~Tacon}},
  \bibinfo{author}{\bibfnamefont{M.}~\bibnamefont{Minola}},
  \bibinfo{author}{\bibfnamefont{S.}~\bibnamefont{Blanco-Canosa}},
  \bibinfo{author}{\bibfnamefont{C.}~\bibnamefont{Mazzoli}},
  \bibinfo{author}{\bibfnamefont{N.~B.} \bibnamefont{Brookes}},
  \bibinfo{author}{\bibfnamefont{G.~M.~D.} \bibnamefont{Luca}},
  \bibinfo{author}{\bibfnamefont{A.}~\bibnamefont{Frano}},
  \bibinfo{author}{\bibfnamefont{D.~G.} \bibnamefont{Hawthorn}},
  \bibinfo{author}{\bibfnamefont{F.}~\bibnamefont{He}}, \bibnamefont{et~al.},
  \bibinfo{journal}{Science} \textbf{\bibinfo{volume}{337}},
  \bibinfo{pages}{821} (\bibinfo{year}{2012}).

\bibitem[{\citenamefont{Chang et~al.}(2012)\citenamefont{Chang, Blackburn,
  Holmes, Christensen, Larsen, Mesot, Liang, Bonn, Hardy, Watenphul
  et~al.}}]{chang12}
\bibinfo{author}{\bibfnamefont{J.}~\bibnamefont{Chang}},
  \bibinfo{author}{\bibfnamefont{E.}~\bibnamefont{Blackburn}},
  \bibinfo{author}{\bibfnamefont{A.~T.} \bibnamefont{Holmes}},
  \bibinfo{author}{\bibfnamefont{N.~B.} \bibnamefont{Christensen}},
  \bibinfo{author}{\bibfnamefont{J.}~\bibnamefont{Larsen}},
  \bibinfo{author}{\bibfnamefont{J.}~\bibnamefont{Mesot}},
  \bibinfo{author}{\bibfnamefont{R.}~\bibnamefont{Liang}},
  \bibinfo{author}{\bibfnamefont{D.~A.} \bibnamefont{Bonn}},
  \bibinfo{author}{\bibfnamefont{W.~N.} \bibnamefont{Hardy}},
  \bibinfo{author}{\bibfnamefont{A.}~\bibnamefont{Watenphul}},
  \bibnamefont{et~al.}, \bibinfo{journal}{Nat. Phys.}
  \textbf{\bibinfo{volume}{8}}, \bibinfo{pages}{871} (\bibinfo{year}{2012}).

\bibitem[{\citenamefont{Achkar et~al.}(2012)\citenamefont{Achkar, Sutarto, Mao,
  He, Frano, Blanco-Canosa, Le~Tacon, Ghiringhelli, Braicovich, Minola
  et~al.}}]{achkar12}
\bibinfo{author}{\bibfnamefont{A.~J.} \bibnamefont{Achkar}},
  \bibinfo{author}{\bibfnamefont{R.}~\bibnamefont{Sutarto}},
  \bibinfo{author}{\bibfnamefont{X.}~\bibnamefont{Mao}},
  \bibinfo{author}{\bibfnamefont{F.}~\bibnamefont{He}},
  \bibinfo{author}{\bibfnamefont{A.}~\bibnamefont{Frano}},
  \bibinfo{author}{\bibfnamefont{S.}~\bibnamefont{Blanco-Canosa}},
  \bibinfo{author}{\bibfnamefont{M.}~\bibnamefont{Le~Tacon}},
  \bibinfo{author}{\bibfnamefont{G.}~\bibnamefont{Ghiringhelli}},
  \bibinfo{author}{\bibfnamefont{L.}~\bibnamefont{Braicovich}},
  \bibinfo{author}{\bibfnamefont{M.}~\bibnamefont{Minola}},
  \bibnamefont{et~al.}, \bibinfo{journal}{Phys. Rev. Lett.}
  \textbf{\bibinfo{volume}{109}}, \bibinfo{pages}{167001}
  (\bibinfo{year}{2012}).

\bibitem[{\citenamefont{Blackburn et~al.}(2013)\citenamefont{Blackburn, Chang,
  H{\"u}cker, Holmes, Christensen, Liang, Bonn, Hardy, R{\"u}tt, Gutowski
  et~al.}}]{blackburn13}
\bibinfo{author}{\bibfnamefont{E.}~\bibnamefont{Blackburn}},
  \bibinfo{author}{\bibfnamefont{J.}~\bibnamefont{Chang}},
  \bibinfo{author}{\bibfnamefont{M.}~\bibnamefont{H{\"u}cker}},
  \bibinfo{author}{\bibfnamefont{A.~T.} \bibnamefont{Holmes}},
  \bibinfo{author}{\bibfnamefont{N.~B.} \bibnamefont{Christensen}},
  \bibinfo{author}{\bibfnamefont{R.}~\bibnamefont{Liang}},
  \bibinfo{author}{\bibfnamefont{D.~A.} \bibnamefont{Bonn}},
  \bibinfo{author}{\bibfnamefont{W.~N.} \bibnamefont{Hardy}},
  \bibinfo{author}{\bibfnamefont{U.}~\bibnamefont{R{\"u}tt}},
  \bibinfo{author}{\bibfnamefont{O.}~\bibnamefont{Gutowski}},
  \bibnamefont{et~al.}, \bibinfo{journal}{Phys. Rev. Lett.}
  \textbf{\bibinfo{volume}{110}}, \bibinfo{pages}{137004}
  (\bibinfo{year}{2013}).

\bibitem[{\citenamefont{Blanco-Canosa et~al.}(2014)\citenamefont{Blanco-Canosa,
  Frano, Schierle, Porras, Loew, Minola, Bluschke, Weschke, Keimer, and
  Le~Tacon}}]{blanco-canosa14}
\bibinfo{author}{\bibfnamefont{S.}~\bibnamefont{Blanco-Canosa}},
  \bibinfo{author}{\bibfnamefont{A.}~\bibnamefont{Frano}},
  \bibinfo{author}{\bibfnamefont{E.}~\bibnamefont{Schierle}},
  \bibinfo{author}{\bibfnamefont{J.}~\bibnamefont{Porras}},
  \bibinfo{author}{\bibfnamefont{T.}~\bibnamefont{Loew}},
  \bibinfo{author}{\bibfnamefont{M.}~\bibnamefont{Minola}},
  \bibinfo{author}{\bibfnamefont{M.}~\bibnamefont{Bluschke}},
  \bibinfo{author}{\bibfnamefont{E.}~\bibnamefont{Weschke}},
  \bibinfo{author}{\bibfnamefont{B.}~\bibnamefont{Keimer}}, \bibnamefont{and}
  \bibinfo{author}{\bibfnamefont{M.} \bibnamefont{Le~Tacon}},
  \bibinfo{journal}{Phys. Rev. B} \textbf{\bibinfo{volume}{90}},
  \bibinfo{pages}{054513} (\bibinfo{year}{2014}).

\bibitem[{\citenamefont{Comin et~al.}(2014)\citenamefont{Comin, Frano, Yee,
  Yoshida, Eisaki, Schierle, Weschke, Sutarto, He, Soumyanarayanan
  et~al.}}]{comin14}
\bibinfo{author}{\bibfnamefont{R.}~\bibnamefont{Comin}},
  \bibinfo{author}{\bibfnamefont{A.}~\bibnamefont{Frano}},
  \bibinfo{author}{\bibfnamefont{M.~M.} \bibnamefont{Yee}},
  \bibinfo{author}{\bibfnamefont{Y.}~\bibnamefont{Yoshida}},
  \bibinfo{author}{\bibfnamefont{H.}~\bibnamefont{Eisaki}},
  \bibinfo{author}{\bibfnamefont{E.}~\bibnamefont{Schierle}},
  \bibinfo{author}{\bibfnamefont{E.}~\bibnamefont{Weschke}},
  \bibinfo{author}{\bibfnamefont{R.}~\bibnamefont{Sutarto}},
  \bibinfo{author}{\bibfnamefont{F.}~\bibnamefont{He}},
  \bibinfo{author}{\bibfnamefont{A.}~\bibnamefont{Soumyanarayanan}},
  \bibnamefont{et~al.}, \bibinfo{journal}{Science}
  \textbf{\bibinfo{volume}{343}}, \bibinfo{pages}{390} (\bibinfo{year}{2014}).

\bibitem[{\citenamefont{da~Silva~Neto et~al.}(2014)\citenamefont{da~Silva~Neto,
  Aynajian, Frano, Comin, Schierle, Weschke, Gyenis, Wen, Schneeloch, Xu
  et~al.}}]{da-silva-neto14}
\bibinfo{author}{\bibfnamefont{E.~H.} \bibnamefont{da~Silva~Neto}},
  \bibinfo{author}{\bibfnamefont{P.}~\bibnamefont{Aynajian}},
  \bibinfo{author}{\bibfnamefont{A.}~\bibnamefont{Frano}},
  \bibinfo{author}{\bibfnamefont{R.}~\bibnamefont{Comin}},
  \bibinfo{author}{\bibfnamefont{E.}~\bibnamefont{Schierle}},
  \bibinfo{author}{\bibfnamefont{E.}~\bibnamefont{Weschke}},
  \bibinfo{author}{\bibfnamefont{A.}~\bibnamefont{Gyenis}},
  \bibinfo{author}{\bibfnamefont{J.}~\bibnamefont{Wen}},
  \bibinfo{author}{\bibfnamefont{J.}~\bibnamefont{Schneeloch}},
  \bibinfo{author}{\bibfnamefont{Z.}~\bibnamefont{Xu}}, \bibnamefont{et~al.},
  \bibinfo{journal}{Science} \textbf{\bibinfo{volume}{343}},
  \bibinfo{pages}{393} (\bibinfo{year}{2014}).

\bibitem[{\citenamefont{Hashimoto et~al.}(2014)\citenamefont{Hashimoto,
  Ghiringhelli, Lee, Dellea, Amorese, Mazzoli, Kummer, Brookes, Moritz, Yoshida
  et~al.}}]{hashimoto14}
\bibinfo{author}{\bibfnamefont{M.}~\bibnamefont{Hashimoto}},
  \bibinfo{author}{\bibfnamefont{G.}~\bibnamefont{Ghiringhelli}},
  \bibinfo{author}{\bibfnamefont{W.-S.} \bibnamefont{Lee}},
  \bibinfo{author}{\bibfnamefont{G.}~\bibnamefont{Dellea}},
  \bibinfo{author}{\bibfnamefont{A.}~\bibnamefont{Amorese}},
  \bibinfo{author}{\bibfnamefont{C.}~\bibnamefont{Mazzoli}},
  \bibinfo{author}{\bibfnamefont{K.}~\bibnamefont{Kummer}},
  \bibinfo{author}{\bibfnamefont{N.~B.} \bibnamefont{Brookes}},
  \bibinfo{author}{\bibfnamefont{B.}~\bibnamefont{Moritz}},
  \bibinfo{author}{\bibfnamefont{Y.}~\bibnamefont{Yoshida}},
  \bibnamefont{et~al.}, \bibinfo{journal}{Phys. Rev. B}
  \textbf{\bibinfo{volume}{89}}, \bibinfo{pages}{220511}
  (\bibinfo{year}{2014}).

\bibitem[{\citenamefont{Tabis et~al.}(2014)\citenamefont{Tabis, Li, Le~Tacon,
  Braicovich, Kreyssig, Minola, Dellea, Weschke, Veit, Ramazanoglu
  et~al.}}]{tabis14}
\bibinfo{author}{\bibfnamefont{W.}~\bibnamefont{Tabis}},
  \bibinfo{author}{\bibfnamefont{Y.}~\bibnamefont{Li}},
  \bibinfo{author}{\bibfnamefont{M.} \bibnamefont{Le~Tacon}},
  \bibinfo{author}{\bibfnamefont{L.}~\bibnamefont{Braicovich}},
  \bibinfo{author}{\bibfnamefont{A.}~\bibnamefont{Kreyssig}},
  \bibinfo{author}{\bibfnamefont{M.}~\bibnamefont{Minola}},
  \bibinfo{author}{\bibfnamefont{G.}~\bibnamefont{Dellea}},
  \bibinfo{author}{\bibfnamefont{E.}~\bibnamefont{Weschke}},
  \bibinfo{author}{\bibfnamefont{M.~J.} \bibnamefont{Veit}},
  \bibinfo{author}{\bibfnamefont{M.}~\bibnamefont{Ramazanoglu}},
  \bibnamefont{et~al.}, \bibinfo{journal}{Nat. Commun.}
  \textbf{\bibinfo{volume}{5}}, \bibinfo{pages}{5875} (\bibinfo{year}{2014}).

\bibitem[{\citenamefont{Kivelson et~al.}(2003)\citenamefont{Kivelson, Bindloss,
  Fradkin, Oganesyan, Tranquada, Kapitulnik, and Howald}}]{kivelson03}
\bibinfo{author}{\bibfnamefont{S.~A.} \bibnamefont{Kivelson}},
  \bibinfo{author}{\bibfnamefont{I.~P.} \bibnamefont{Bindloss}},
  \bibinfo{author}{\bibfnamefont{E.}~\bibnamefont{Fradkin}},
  \bibinfo{author}{\bibfnamefont{V.}~\bibnamefont{Oganesyan}},
  \bibinfo{author}{\bibfnamefont{J.~M.} \bibnamefont{Tranquada}},
  \bibinfo{author}{\bibfnamefont{A.}~\bibnamefont{Kapitulnik}},
  \bibnamefont{and} \bibinfo{author}{\bibfnamefont{C.}~\bibnamefont{Howald}},
  \bibinfo{journal}{Rev. Mod. Phys.} \textbf{\bibinfo{volume}{75}},
  \bibinfo{pages}{1201} (\bibinfo{year}{2003}).

\bibitem[{\citenamefont{da~Silva~Neto et~al.}(2015)\citenamefont{da~Silva~Neto,
  Comin, He, Sutarto, Jiang, Greene, Sawatzky, and
  Damascelli}}]{da-silva-neto15}
\bibinfo{author}{\bibfnamefont{E.~H.} \bibnamefont{da~Silva~Neto}},
  \bibinfo{author}{\bibfnamefont{R.}~\bibnamefont{Comin}},
  \bibinfo{author}{\bibfnamefont{F.}~\bibnamefont{He}},
  \bibinfo{author}{\bibfnamefont{R.}~\bibnamefont{Sutarto}},
  \bibinfo{author}{\bibfnamefont{Y.}~\bibnamefont{Jiang}},
  \bibinfo{author}{\bibfnamefont{R.~L.} \bibnamefont{Greene}},
  \bibinfo{author}{\bibfnamefont{G.~A.} \bibnamefont{Sawatzky}},
  \bibnamefont{and}
  \bibinfo{author}{\bibfnamefont{A.}~\bibnamefont{Damascelli}},
  \bibinfo{journal}{Science} \textbf{\bibinfo{volume}{347}},
  \bibinfo{pages}{282} (\bibinfo{year}{2015}).

\bibitem[{mis({\natexlab{a}})}]{misc-PGelectron}
\bibinfo{note}{A pseudogap was reported in a state above the antiferromagnetic
  state in electron-doped cuprates \cite{onose01}, but the pseudogap similar to
  the hole-doped case, i.e., a gap-like feature above the onset temperature of
  superconducting instability is missing or at least very weak.}

\bibitem[{\citenamefont{Yamase et~al.}(2015)\citenamefont{Yamase, Bejas, and
  Greco}}]{yamase15b}
\bibinfo{author}{\bibfnamefont{H.}~\bibnamefont{Yamase}},
  \bibinfo{author}{\bibfnamefont{M.}~\bibnamefont{Bejas}}, \bibnamefont{and}
  \bibinfo{author}{\bibfnamefont{A.}~\bibnamefont{Greco}},
  \bibinfo{journal}{Europhys. Lett.} \textbf{\bibinfo{volume}{111}},
  \bibinfo{pages}{57005} (\bibinfo{year}{2015}).

\bibitem[{\citenamefont{Ishii et~al.}(2014)\citenamefont{Ishii, Fujita, Sasaki,
  Minola, Dellea, Mazzoli, Kummer, Ghiringhelli, Braicovich, Tohyama
  et~al.}}]{ishii14}
\bibinfo{author}{\bibfnamefont{K.}~\bibnamefont{Ishii}},
  \bibinfo{author}{\bibfnamefont{M.}~\bibnamefont{Fujita}},
  \bibinfo{author}{\bibfnamefont{T.}~\bibnamefont{Sasaki}},
  \bibinfo{author}{\bibfnamefont{M.}~\bibnamefont{Minola}},
  \bibinfo{author}{\bibfnamefont{G.}~\bibnamefont{Dellea}},
  \bibinfo{author}{\bibfnamefont{C.}~\bibnamefont{Mazzoli}},
  \bibinfo{author}{\bibfnamefont{K.}~\bibnamefont{Kummer}},
  \bibinfo{author}{\bibfnamefont{G.}~\bibnamefont{Ghiringhelli}},
  \bibinfo{author}{\bibfnamefont{L.}~\bibnamefont{Braicovich}},
  \bibinfo{author}{\bibfnamefont{T.}~\bibnamefont{Tohyama}},
  \bibnamefont{et~al.}, \bibinfo{journal}{Nat. Commun.}
  \textbf{\bibinfo{volume}{5}}, \bibinfo{pages}{3714} (\bibinfo{year}{2014}).

\bibitem[{\citenamefont{Lee et~al.}(2014)\citenamefont{Lee, Lee, Nowadnick,
  Gerber, Tabis, Huang, Strocov, Motoyama, Yu, Moritz et~al.}}]{wslee14}
\bibinfo{author}{\bibfnamefont{W.~S.} \bibnamefont{Lee}},
  \bibinfo{author}{\bibfnamefont{J.~J.} \bibnamefont{Lee}},
  \bibinfo{author}{\bibfnamefont{E.~A.} \bibnamefont{Nowadnick}},
  \bibinfo{author}{\bibfnamefont{S.}~\bibnamefont{Gerber}},
  \bibinfo{author}{\bibfnamefont{W.}~\bibnamefont{Tabis}},
  \bibinfo{author}{\bibfnamefont{S.~W.} \bibnamefont{Huang}},
  \bibinfo{author}{\bibfnamefont{V.~N.} \bibnamefont{Strocov}},
  \bibinfo{author}{\bibfnamefont{E.~M.} \bibnamefont{Motoyama}},
  \bibinfo{author}{\bibfnamefont{G.}~\bibnamefont{Yu}},
  \bibinfo{author}{\bibfnamefont{B.}~\bibnamefont{Moritz}},
  \bibnamefont{et~al.}, \bibinfo{journal}{Nat. Phys.}
  \textbf{\bibinfo{volume}{10}}, \bibinfo{pages}{883} (\bibinfo{year}{2014}).

\bibitem[{\citenamefont{Singley et~al.}(2001)\citenamefont{Singley, Basov,
  Kurahashi, Uefuji, and Yamada}}]{singley01}
\bibinfo{author}{\bibfnamefont{E.~J.} \bibnamefont{Singley}},
  \bibinfo{author}{\bibfnamefont{D.~N.} \bibnamefont{Basov}},
  \bibinfo{author}{\bibfnamefont{K.}~\bibnamefont{Kurahashi}},
  \bibinfo{author}{\bibfnamefont{T.}~\bibnamefont{Uefuji}}, \bibnamefont{and}
  \bibinfo{author}{\bibfnamefont{K.}~\bibnamefont{Yamada}},
  \bibinfo{journal}{Phys.\ Rev.\ B} \textbf{\bibinfo{volume}{64}},
  \bibinfo{pages}{224503} (\bibinfo{year}{2001}).

\bibitem[{\citenamefont{N\"{u}cker et~al.}(1989)\citenamefont{N\"{u}cker,
  Romberg, Nakai, Scheerer, Fink, Yan, and Zhao}}]{nuecker89}
\bibinfo{author}{\bibfnamefont{N.}~\bibnamefont{N\"{u}cker}},
  \bibinfo{author}{\bibfnamefont{H.}~\bibnamefont{Romberg}},
  \bibinfo{author}{\bibfnamefont{S.}~\bibnamefont{Nakai}},
  \bibinfo{author}{\bibfnamefont{B.}~\bibnamefont{Scheerer}},
  \bibinfo{author}{\bibfnamefont{J.}~\bibnamefont{Fink}},
  \bibinfo{author}{\bibfnamefont{Y.~F.} \bibnamefont{Yan}}, \bibnamefont{and}
  \bibinfo{author}{\bibfnamefont{Z.~X.} \bibnamefont{Zhao}},
  \bibinfo{journal}{Phys. Rev. B} \textbf{\bibinfo{volume}{39}},
  \bibinfo{pages}{12 379} (\bibinfo{year}{1989}).

\bibitem[{\citenamefont{Romberg et~al.}(1990)\citenamefont{Romberg, N\"{u}cker,
  Fink, Wolf, Xi, Koch, Geserich, D\"{u}rrler, Assmus, and
  Gegenheimer}}]{romberg90}
\bibinfo{author}{\bibfnamefont{H.}~\bibnamefont{Romberg}},
  \bibinfo{author}{\bibfnamefont{N.}~\bibnamefont{N\"{u}cker}},
  \bibinfo{author}{\bibfnamefont{J.}~\bibnamefont{Fink}},
  \bibinfo{author}{\bibfnamefont{T.}~\bibnamefont{Wolf}},
  \bibinfo{author}{\bibfnamefont{X.~X.} \bibnamefont{Xi}},
  \bibinfo{author}{\bibfnamefont{B.}~\bibnamefont{Koch}},
  \bibinfo{author}{\bibfnamefont{H.~P.} \bibnamefont{Geserich}},
  \bibinfo{author}{\bibfnamefont{M.}~\bibnamefont{D\"{u}rrler}},
  \bibinfo{author}{\bibfnamefont{W.}~\bibnamefont{Assmus}}, \bibnamefont{and}
  \bibinfo{author}{\bibfnamefont{B.}~\bibnamefont{Gegenheimer}},
  \bibinfo{journal}{Z. Phys. B} \textbf{\bibinfo{volume}{78}},
  \bibinfo{pages}{367} (\bibinfo{year}{1990}).

\bibitem[{\citenamefont{Kresin and Morawitz}(1988)}]{kresin88}
\bibinfo{author}{\bibfnamefont{V.~Z.} \bibnamefont{Kresin}} \bibnamefont{and}
  \bibinfo{author}{\bibfnamefont{H.}~\bibnamefont{Morawitz}},
  \bibinfo{journal}{Phys. Rev. B} \textbf{\bibinfo{volume}{37}},
  \bibinfo{pages}{7854} (\bibinfo{year}{1988}).

\bibitem[{\citenamefont{Bill et~al.}(2003)\citenamefont{Bill, Morawitz, and
  Kresin}}]{bill03}
\bibinfo{author}{\bibfnamefont{A.}~\bibnamefont{Bill}},
  \bibinfo{author}{\bibfnamefont{H.}~\bibnamefont{Morawitz}}, \bibnamefont{and}
  \bibinfo{author}{\bibfnamefont{V.~Z.} \bibnamefont{Kresin}},
  \bibinfo{journal}{Phys. Rev. B} \textbf{\bibinfo{volume}{68}},
  \bibinfo{pages}{144519} (\bibinfo{year}{2003}).

\bibitem[{\citenamefont{van Loon et~al.}(2014)\citenamefont{van Loon,
  Hafermann, Lichtenstein, Rubtsov, and Katsnelson}}]{vanloon14}
\bibinfo{author}{\bibfnamefont{E.~G. C.~P.} \bibnamefont{van Loon}},
  \bibinfo{author}{\bibfnamefont{H.}~\bibnamefont{Hafermann}},
  \bibinfo{author}{\bibfnamefont{A.~I.} \bibnamefont{Lichtenstein}},
  \bibinfo{author}{\bibfnamefont{A.~N.} \bibnamefont{Rubtsov}},
  \bibnamefont{and} \bibinfo{author}{\bibfnamefont{M.~I.}
  \bibnamefont{Katsnelson}}, \bibinfo{journal}{Phys. Rev. Lett.}
  \textbf{\bibinfo{volume}{113}}, \bibinfo{pages}{246407}
  (\bibinfo{year}{2014}).

\bibitem[{\citenamefont{Markiewicz et~al.}(2008)\citenamefont{Markiewicz,
  Hasan, and Bansil}}]{markiewicz08}
\bibinfo{author}{\bibfnamefont{R.~S.} \bibnamefont{Markiewicz}},
  \bibinfo{author}{\bibfnamefont{M.~Z.} \bibnamefont{Hasan}}, \bibnamefont{and}
  \bibinfo{author}{\bibfnamefont{A.}~\bibnamefont{Bansil}},
  \bibinfo{journal}{Phys. Rev. B} \textbf{\bibinfo{volume}{77}},
  \bibinfo{pages}{094518} (\bibinfo{year}{2008}).

\bibitem[{\citenamefont{{A. Foussats and A. Greco}}(2004)}]{foussats04}
\bibinfo{author}{\bibnamefont{{A. Foussats and A. Greco}}},
  \bibinfo{journal}{Phys.\ Rev.\ B} \textbf{\bibinfo{volume}{70}},
  \bibinfo{pages}{205123} (\bibinfo{year}{2004}).

\bibitem[{\citenamefont{Thio et~al.}(1988)\citenamefont{Thio, Thurston, Preyer,
  Picone, Kastner, Jenssen, Gabbe, Chen, Birgeneau, and Aharony}}]{thio88}
\bibinfo{author}{\bibfnamefont{T.}~\bibnamefont{Thio}},
  \bibinfo{author}{\bibfnamefont{T.~R.} \bibnamefont{Thurston}},
  \bibinfo{author}{\bibfnamefont{N.~W.} \bibnamefont{Preyer}},
  \bibinfo{author}{\bibfnamefont{P.~J.} \bibnamefont{Picone}},
  \bibinfo{author}{\bibfnamefont{M.~A.} \bibnamefont{Kastner}},
  \bibinfo{author}{\bibfnamefont{H.~P.} \bibnamefont{Jenssen}},
  \bibinfo{author}{\bibfnamefont{D.~R.} \bibnamefont{Gabbe}},
  \bibinfo{author}{\bibfnamefont{C.~Y.} \bibnamefont{Chen}},
  \bibinfo{author}{\bibfnamefont{R.~J.} \bibnamefont{Birgeneau}},
  \bibnamefont{and} \bibinfo{author}{\bibfnamefont{A.}~\bibnamefont{Aharony}},
  \bibinfo{journal}{Phys. Rev. B} \textbf{\bibinfo{volume}{38}},
  \bibinfo{pages}{905} (\bibinfo{year}{1988}).

\bibitem[{\citenamefont{Becca et~al.}(1996)\citenamefont{Becca, Tarquini,
  Grilli, and Di~Castro}}]{becca96}
\bibinfo{author}{\bibfnamefont{F.}~\bibnamefont{Becca}},
  \bibinfo{author}{\bibfnamefont{M.}~\bibnamefont{Tarquini}},
  \bibinfo{author}{\bibfnamefont{M.}~\bibnamefont{Grilli}}, \bibnamefont{and}
  \bibinfo{author}{\bibfnamefont{C.} \bibnamefont{Di~Castro}},
  \bibinfo{journal}{Phys.\ Rev.\ B} \textbf{\bibinfo{volume}{54}},
  \bibinfo{pages}{12443} (\bibinfo{year}{1996}).

\bibitem[{\citenamefont{Bejas et~al.}(2012)\citenamefont{Bejas, Greco, and
  Yamase}}]{bejas12}
\bibinfo{author}{\bibfnamefont{M.}~\bibnamefont{Bejas}},
  \bibinfo{author}{\bibfnamefont{A.}~\bibnamefont{Greco}}, \bibnamefont{and}
  \bibinfo{author}{\bibfnamefont{H.}~\bibnamefont{Yamase}},
  \bibinfo{journal}{Phys.\ Rev.\ B} \textbf{\bibinfo{volume}{86}},
  \bibinfo{pages}{224509} (\bibinfo{year}{2012}).

\bibitem[{\citenamefont{Bejas et~al.}(2014)\citenamefont{Bejas, Greco, and
  Yamase}}]{bejas14}
\bibinfo{author}{\bibfnamefont{M.}~\bibnamefont{Bejas}},
  \bibinfo{author}{\bibfnamefont{A.}~\bibnamefont{Greco}}, \bibnamefont{and}
  \bibinfo{author}{\bibfnamefont{H.}~\bibnamefont{Yamase}},
  \bibinfo{journal}{New J. Phys.} \textbf{\bibinfo{volume}{16}},
  \bibinfo{pages}{123002} (\bibinfo{year}{2014}).

\bibitem[{\citenamefont{Andersen et~al.}(1995)\citenamefont{Andersen,
  Lichtenstein, Jepsen, and Paulsen}}]{andersen95}
\bibinfo{author}{\bibfnamefont{O.~K.} \bibnamefont{Andersen}},
  \bibinfo{author}{\bibfnamefont{A.~I.} \bibnamefont{Lichtenstein}},
  \bibinfo{author}{\bibfnamefont{O.}~\bibnamefont{Jepsen}}, \bibnamefont{and}
  \bibinfo{author}{\bibfnamefont{F.}~\bibnamefont{Paulsen}},
  \bibinfo{journal}{J.\ Phys.\ Chem.\ Solids} \textbf{\bibinfo{volume}{56}},
  \bibinfo{pages}{1573} (\bibinfo{year}{1995}).

\bibitem[{\citenamefont{Hoang and Thalmeier}(2002)}]{hoang02}
\bibinfo{author}{\bibfnamefont{A.~T.} \bibnamefont{Hoang}} \bibnamefont{and}
  \bibinfo{author}{\bibfnamefont{P.}~\bibnamefont{Thalmeier}},
  \bibinfo{journal}{J. Phys.: Condens. Matter} \textbf{\bibinfo{volume}{14}},
  \bibinfo{pages}{6639} (\bibinfo{year}{2002}).

\bibitem[{\citenamefont{{J. Merino, A. Greco, R.~H. McKenzie, and M.
  Calandra}}(2003)}]{merino03}
\bibinfo{author}{\bibnamefont{{J. Merino, A. Greco, R.~H. McKenzie, and M.
  Calandra}}}, \bibinfo{journal}{Phys.\ Rev.\ B} \textbf{\bibinfo{volume}{68}},
  \bibinfo{pages}{245121} (\bibinfo{year}{2003}).

\bibitem[{mis({\natexlab{b}})}]{misc-d}
\bibinfo{note}{Although CuO$_2$ planes shift by $(\frac{1}{2}, \frac{1}{2},
  \frac{1}{2})$ in NCCO, we model the actual system by neglecting such a shift
  for simplicity. Our interlayer distance $d$ is thus given by a half of the
  $c$-axis lattice constant.}

\bibitem[{\citenamefont{Prelov\v{s}ek and Horsch}(1999)}]{prelovsek99}
\bibinfo{author}{\bibfnamefont{P.}~\bibnamefont{Prelov\v{s}ek}}
  \bibnamefont{and} \bibinfo{author}{\bibfnamefont{P.}~\bibnamefont{Horsch}},
  \bibinfo{journal}{Phys. Rev. B} \textbf{\bibinfo{volume}{60}},
  \bibinfo{pages}{R3735} (\bibinfo{year}{1999}).

\bibitem[{\citenamefont{Timusk and Tanner}(1989)}]{timusk89}
\bibinfo{author}{\bibfnamefont{T.}~\bibnamefont{Timusk}} \bibnamefont{and}
  \bibinfo{author}{\bibfnamefont{D.}~\bibnamefont{Tanner}},
  \emph{\bibinfo{title}{Infrared properties of high-Tc superconductors}}
  (\bibinfo{publisher}{Word Scientific, Singapure}, \bibinfo{year}{1989}).

\bibitem[{\citenamefont{Mahan}(1990)}]{mahan}
\bibinfo{author}{\bibfnamefont{G.~D.} \bibnamefont{Mahan}},
  \emph{\bibinfo{title}{Many-Particle Physics}} (\bibinfo{publisher}{Plenum,
New York}, \bibinfo{year}{1990}), \bibinfo{edition}{2nd} ed.

\bibitem[{\citenamefont{Hybertsen et~al.}(1990)\citenamefont{Hybertsen,
  Stechel, Schluter, and Jennison}}]{hybertsen90}
\bibinfo{author}{\bibfnamefont{M.~S.} \bibnamefont{Hybertsen}},
  \bibinfo{author}{\bibfnamefont{E.~B.} \bibnamefont{Stechel}},
  \bibinfo{author}{\bibfnamefont{M.}~\bibnamefont{Schluter}}, \bibnamefont{and}
  \bibinfo{author}{\bibfnamefont{D.~R.} \bibnamefont{Jennison}},
  \bibinfo{journal}{Phys.\ Rev.\ B} \textbf{\bibinfo{volume}{41}},
  \bibinfo{pages}{11068} (\bibinfo{year}{1990}).

\bibitem[{mis({\natexlab{c}})}]{misc-factor2}
\bibinfo{note}{A factor of $1/2$ here comes from a large-$N$ formalism where
  $t$ is scaled by $1/N$. We may assume $N=2$ in comparison with experiments.}

\bibitem[{\citenamefont{Hwang et~al.}(2007)\citenamefont{Hwang, Timusk, and
  Gu}}]{hwang07}
\bibinfo{author}{\bibfnamefont{J.}~\bibnamefont{Hwang}},
  \bibinfo{author}{\bibfnamefont{T.}~\bibnamefont{Timusk}}, \bibnamefont{and}
  \bibinfo{author}{\bibfnamefont{G.}~\bibnamefont{Gu}}, \bibinfo{journal}{J.
  Phys.: Condens. Matter} \textbf{\bibinfo{volume}{19}},
  \bibinfo{pages}{125208} (\bibinfo{year}{2007}).

\bibitem[{mis({\natexlab{d}})}]{misc-ishii}
\bibinfo{note}{ K. Ishii (private communication)}.

\bibitem[{\citenamefont{{Y. Onose and Y. Taguchi and K. Ishizaka and Y.
  Tokura}}(2001)}]{onose01}
\bibinfo{author}{\bibnamefont{{Y. Onose and Y. Taguchi and K. Ishizaka and Y.
  Tokura}}}, \bibinfo{journal}{Phys.\ Rev.\ Lett.}
  \textbf{\bibinfo{volume}{87}}, \bibinfo{pages}{217001}
  (\bibinfo{year}{2001}).

\end{thebibliography}

\end{document}